\documentclass[a4paper]{jpconf}
\usepackage{graphicx}
\begin{document}
\title{Wave packet methods in cavity QED}

\author{Jonas Larson}

\address{ICFO-Institut de Ci\`{e}ncies Fot\`{o}niques, E-08860 Castelldefels, Barcelona, Spain}

\ead{jonas.larson@icfo.es}

\begin{abstract}
The Jaynes-Cummings model, with and without the rotating wave
approximation, is expressed in the conjugate variable representation
and solved numerically by wave packet propagation. Both cases are
then cast into systems of two coupled harmonic oscillators,
reminiscent of coupled bound electronic potential curves of diatomic
molecules. Using the knowledge of such models, this approach of the
problem gives new insight of the dynamics. The effect of the
rotating wave approximation is discussed. The collapse-revival
phenomenon is especially analyzed in a non-standard manner.
Extensions of the method is briefly mentioned in terms of a
three-level atom and the Dicke model.

\end{abstract}

\section{Introduction}
The main concepts of a wave packet approach to cavity quantum
electrodynamics (QED) where put forward in \cite{oldwine}. The idea
is to formulate the model Hamiltonian, either the {\it
Jaynes-Cummings} (JC) model or the {\it Rabi} model \cite{jc,rabi},
in terms of the quadrature operators for the field rather than the
commonly used boson ladder operators. The quadrature operators obey
the regular conjugate commutator relations like position and
momentum of a particle. As such, the problem can be viewed as a wave
packet evolving on two coupled and bound {\it potential curves}
(arising due to the two-level structure of the considered atom). The
state of the system is, of course, embedded in the wave packet,
where roughly speaking the potential curves determines the internal
state of the two-level atom and the vibrational states of the wave
packet correspond to the field mode state. Thus, once the wave
packet is obtained any quantity is easily calculated. The JC and
Rabi models are related through the {\it rotating wave
approximation} (RWA) in which rapidly oscillating terms are
neglected in the Rabi model to give the JC one. Correspondingly, the
JC model is exactly solvable and the physical quantities of interest
are in general represented by infinite sums deriving from the
quantized nature of the cavity field. These analytical results are
valid only in the regime where the RWA is justified, and beyond
these the Rabi model must be considered. In the conjugate variable
representation, both models exhibit an {\it avoided crossing}
between the two potential curves, crucially affecting their
dynamics. Surprisingly, especially around the avoided crossing the
solvable JC model shows a non-intuitive wave packet dynamics, while
the Rabi model display a more expected evolution.

In this paper we extend some of the results of \cite{oldwine} and
describe how the method can be generalized to multi-level systems.
Explicitly, we deepen the analysis about the phenomenon of
collapse-revivals and discuss the differences in the evolution
between the two models. The approach to collapse-revivals in the JC
and Rabi models is somewhat different from the one typically used
for wave packets evolution in molecular and chemical physics. This
is thoroughly considered and the relation between the two viewpoints
is sorted out. In particular, what is characterized by the {\it
revival time} $T_{rev}$ for the JC model is labeled the {\it
classical period} in wave packet dynamics. Thus, the time scale set
by $T_{rev}$ can become very long compared to what one would expect
from the classical period, which comes about due to the internal
two-level structure of the model. Unlike \cite{oldwine}, this paper
briefly considers multi-level systems, that is when the number of
internal states exceeds 2. We first consider a simple example of a
three-level $\Lambda$-atom coupled to a single cavity mode, which
serves as a prototype of how to extend the theory. More interesting
is the following system of $N$ two-level atoms coupled to a
quantized field, namely the {\it Dicke} model \cite{dicke}. Here,
the proper basis for the atomic subsystem is a collective one
reminiscent of angular momentum states. Within this basis, the
adiabatic diagonalization is straightforward and the $2^N$ potential
curves are regained. An alternative approach is to use the
Holstein-Primakoff representation \cite{hp}, in which the $2^N$
atomic (fermionic) degrees of freedom is replaced by a single
bosonic degree of freedom. In such case, the coupled potential
curves are represented by one potential surface in 2-D.

We proceed as follows. In the next section \ref{sec2} we introduce
the model systems, the relation between the two models \ref{ssec2a}
is discussed and we give both models in the conjugate variable
representation in \ref{ssec2b}. In this subsection we also mention
how one may derive a semi-classical model that describes the
population transfer between the two potential curves as the wave
packet traverses the crossing. A deeper insight of the coupled
dynamics is gained from the adiabatic representation, presented in
\ref{ssec2c}. The following section \ref{sec3} considers the
non-intuitive evolution of the JC model compared to the Rabi model,
while section \ref{sec4} in detail studies the collapse-revival
phenomenon, emphasizing on parallels between these two models and
other wave packet systems. Section \ref{sec5} is devoted to
generalizations of the method to multi-level systems. Firstly, in
\ref{ssec5a} the idea is sketched using a simple three-level system,
and in \ref{ssec5b} we consider the Dicke model. Finally, section
\ref{sec6} gives a summary and outlooks.

\section{The model system}\label{sec2}
\subsection{Relation between the Rabi and the Jaynes-Cummings
models}\label{ssec2a} In most cavity QED experiments, their
characteristics is well described by the Jaynes-Cummings model, in
which single modes and atomic transitions are isolated and
coherently coupled. This is achievable due to the strong atom-field
coupling and the use of high-$Q$ cavities and long-lived atomic
states, such that losses can be discarded over typical interaction
periods. Within the dipole approximation, a microscopical derivation
leads to the {\it Rabi model} defined by the Hamiltonian
\begin{equation}\label{hrabi}
H'_{Rabi}=\hbar\omega \left(a^\dagger
a+\frac{1}{2}\right)+\frac{\hbar\Omega'}{2}\sigma_z+\hbar
g'_0\left(\sigma^++\sigma^-\right)\left(a^\dagger+a\right).
\end{equation}
Here, $a^\dagger$ ($a$) is the creation (annihilation) operator for
the field mode; $a|n\rangle=\sqrt{n+1}|n+1\rangle$
($a|n\rangle=\sqrt{n}|n-1\rangle$), the sigma operators are the
standard Pauli matrices acting on the two-level atom;
$\sigma_z|\pm\rangle=\pm|\pm\rangle$,
$\sigma^\pm|\pm\rangle=|\mp\rangle$, $\sigma_x=\sigma^++\sigma^-$
and $\sigma_y=-i\left(\sigma^+-\sigma^-\right)$, $\omega$ ($\Omega$)
the field (atomic transition) frequency and $g'_0$ is the effective
atom-field coupling. Before proceeding, we define characteristic
time and energy scales by $\omega^{-1}$ and $\hbar\omega$
respectively, and hence introduce the dimensionless variables
$H_{Rabi}=H'_{Rabi}/\hbar\omega$, $\Omega=\Omega'/\omega$ and
$g_0=g'_0/\omega$. The interaction in (\ref{hrabi}) is built up from
four terms; $\sigma^+a^\dagger$ and $\sigma^-a$ corresponding to
simultaneous excitation/deaxcitation of the atom and the field, and
$\sigma^+a$ and $a^\dagger\sigma^-$ originating from excitation of
the atom by absorption of one photon and vice versa  While the first
terms are appropriately called {\it non-energy conserving terms},
the latter are termed {\it energy conserving terms}. In a rotating
frame with respect to the first two terms of (\ref{hrabi}), the
interaction constitutes precess with either the frequency
$|\Omega+1|$ or $|\Omega-1|$, and provided
\begin{equation}\label{rwacon}
|\Omega-1|\ll|\Omega+1|
\end{equation}
the rapidly oscillating terms may be left out. In this RWA limit,
the Rabi Hamiltonian relaxes to the acclaimed Jaynes-Cummings model
\begin{equation}\label{hjc0}
H_{JC}=\left(a^\dagger
a+\frac{1}{2}\right)+\frac{\hbar\Omega}{2}\sigma_z+\hbar
g_0\left(\sigma^+a+\sigma^-a^\dagger\right).
\end{equation}
The applicability of the RWA is not solely determined by the
requirement (\ref{rwacon}), but also on the ratio $g_0/\Omega$. For
large values of $g_0/\Omega$, the typical time scale of the
interaction exceeds the one of free atomic evolution and the
time-accumulated error arising from neglecting the non-energy
conserving terms becomes important. Thus, apart from the condition
(\ref{rwacon}) between the atomic transition and field frequencies,
the validity of the RWA necessitates that $g_0<\Omega$ \cite{com1}.

Given within the RWA, the number of excitations $N=a^\dagger
a+\frac{1}{2}\sigma_z$ is a conserved quantity, and taking this
symmetry into account the JC model is readily solvable with
eigenstates
\begin{equation}\label{eigs}
\begin{array}{l}
\displaystyle{|E_n\rangle_+=\cos\left(\frac{\theta}{2}\right)|+,n-1\rangle+\sin\left(\frac{\theta}{2}\right)|-,n\rangle,}\\ \\
\displaystyle{|E_n\rangle_-=\sin\left(\frac{\theta}{2}\right)|+,n-1\rangle-\cos\left(\frac{\theta}{2}\right)|-,n\rangle,}
\end{array}
\end{equation}
where
\begin{equation}
\tan(2\theta)=\frac{2g_0\sqrt{n}}{\Delta}
\end{equation}
and $\Delta=\Omega-1$. The corresponding eigenvalues read
\begin{equation}
E_n^\pm=\left(n+\frac{1}{2}\right)\pm\sqrt{\frac{\Delta^2}{4}+g_0^2n},
\end{equation}
together with the uncoupled ground state $|E_0\rangle=|-,0\rangle$
with $E_0=-\Omega/2$.

\subsection{Conjugate variable representation}\label{ssec2b}
The algebraic approach applied to the models presented in the
previous section is, in many cases, preferable to other methods. In
terms of the Rabi Hamiltonian, no exact analytical solutions exist
and approximate concepts are developed to find the required
quantities in certain parameter regimes \cite{oldwine}. For example,
this enables for perturbation expansions or truncation of continued
fraction solutions. Non the less, the validity of the results are
restricted to specific parameters, and to go beyond these one must
consider numerical approaches. The procedure used here adopts the
$x$-representation in which the boson operators define the conjugate
variables
\begin{equation}
p=i\frac{1}{\sqrt{2}}\left(a^\dagger-a\right),\hspace{1cm}x=\frac{1}{\sqrt{2}}\left(a^\dagger+a\right)
\end{equation}
obeying $[x,p]=i$. In this representation, the Hamiltonians
(\ref{hrabi}) and (\ref{hjc0}) become
\begin{equation}\label{conrabi}
\displaystyle{H_{Rabi}=\frac{p^2}{2}+\frac{x^2}{2}+\left[\begin{array}{cc}\displaystyle{\frac{\Omega}{2}}
& g_0\sqrt{2}x \\
g_0\sqrt{2}x &
-\displaystyle{\frac{\Omega}{2}}\end{array}\right]},
\end{equation}
\begin{equation}\label{conjc}
\displaystyle{H_{JC}=\frac{p^2}{2}+\frac{x^2}{2}+\left[\begin{array}{cc}\displaystyle{\frac{\Omega}{2}}
&
\displaystyle{\frac{g_0}{\sqrt{2}}\left(x+ip\right)}
\\
\displaystyle{\frac{g_0}{\sqrt{2}}\left(x-ip\right)}
& -\displaystyle{\frac{\Omega}{2}}\end{array}\right]}.
\end{equation}
It is convenient to unitarily rotate $H_{Rabi}$ and $H_{JC}$ by
$U=\frac{1}{\sqrt{2}}(\sigma_x+\sigma_z)$ which gives the
transformed Hamiltonians
\begin{equation}
\tilde{H}_{Rabi}=\frac{p^2}{2}+\left[\begin{array}{cc}V_h\left(x+\sqrt{2}g_0\right)
&  \displaystyle{\frac{\Omega}{2}} \\
\displaystyle{\frac{\Omega}{2}} &
V_h\left(Q-\sqrt{2}g_0\right)\end{array}\right]-g_0^2,
\end{equation}
\begin{equation}
\tilde{H}_{JC}=\frac{p^2}{2}+\left[\begin{array}{cc}V_h\left(x+\displaystyle{\frac{g_0}{\sqrt{2}}}\right)
&  \displaystyle{\frac{\Omega}{2}-i\frac{g_0}{\sqrt{2}}p} \\
\displaystyle{\frac{\Omega}{2}+i{\frac{g_0}{\sqrt{2}}p}}
&
V_h\left(x-\displaystyle{\frac{g_0}{\sqrt{2}}}\right)\end{array}\right]-\frac{g_0^2}{4},
\end{equation}
where $V_h(x)=x^2/2$. In the current nomenclature both models are
identified as two displaced coupled harmonic oscillators, in which,
however, the amount of displacement and the character of the
couplings are emphatically different. For $x=0$ the two potential
curves possess an avoided crossing and the dynamics around such
degenrate points have been thoroughly analyzed in the studies of
excited diatomic molecules \cite{barry}. Typically, the crucial part
of the evolution occurs when the wave packet transverses an avoided
crossing. To a good approximation the potentials can be linearized
in the vicinity of the crossing. Considering the wave packet as a
classical point particle following the classical equations of
motion, the population transfer between the two levels while passing
the crossing in the Rabi model can be estimated by the {\it
Landau-Zener formula} \cite{oldwine,LZ1,LZ2}
\begin{equation}\label{lzform}
P_{LZ}=1-\exp\left(-\frac{\sqrt{2}\pi\Omega^2}{8g_0v}\right).
\end{equation}
Here $v$ is the classical velocity of the wave packet at the
crossing, which, of course, depends on the initial state. Equation
(\ref{lzform}) directly gives some indicatives of the system
behaviours; for a large velocity $v$ only a small fraction is
transferred to the opposite state, and the same holds for a modest
parameter $\Omega$ which couples the two levels. In order for the
wave packet to bisect the crossing region, its initial mean position
$x_i$ must satisfy $x_i<0$ or $x_i^2>2g_0^2$ for an initial state
starting out ``on'' the right shifted oscillator or $x_i>0$ or
$x_i^2<-2g_0^2$ for the left oscillator. For such a state, starting
out on a single potential curve, a classical estimate of the
velocity at the crossing gives $v\approx\sqrt{x_i^2-g_0^2/2}$. The
evolutions is said to be {\it adiabatic} if $P_{LZ}\approx1$, {\it
diabatic} in the opposite limit and {\it mesobatic} in the
intermediate regime. The same arguments applied to the JC model is
considerably less direct as the two potential curves are coupled by
a ``momentum'' dependent term. This term is the background to the
aberrance between the two models in a most unexpected way. For
$|\Omega|\gg g_0\sqrt{\bar{n}}$, where $\bar{n}$ is the average
number of photons (directly related to the velocity $v$), it follows
that the equations can be decoupled, which corresponds to adiabatic
elimination of the two atomic internal states. This will be
confirmed in the next subsection.

\subsection{Adiabatic diagonalization}\label{ssec2c}
The previous section introduced the idea of potential curves
associated with the Rabi and JC models, and the concept of
adiabaticity became somewhat clear in this picture. Here we will
stress this even further in order to gain a deeper understanding of
the dynamics. As the JC model is exactly solvable we only carry out
the analysis for the Rabi model.

\begin{figure}[ht]
\begin{center}
\includegraphics[width=12cm]{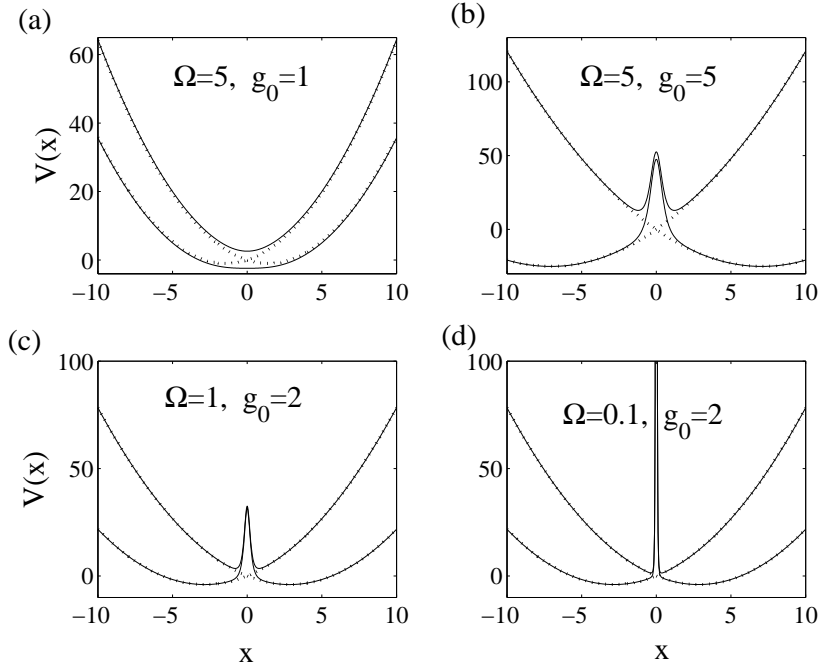}
\end{center}
\caption{\label{fig1} This figure displays different examples of the
adiabatic potentials $V_{ad}^\pm(x)$ (solid lines) and diabatic
potentials $V_d^\pm(x)$ (dotted lines). }
\end{figure}

The basis in which the Hamiltonian (\ref{conrabi}) is written might
not be the most ``optimal'' one in the sense of decoupling the
internal states. In most cases, the {\it potential matrix}
describing the coupled dynamics has a smooth $x$-dependence, such
that the characteristic length of derivative terms of the coupling
elements decrease with the order of derivatives. A systematic way to
remove the low-derivative terms from the off diagonals is by the
{\it adiabatic diagonalization}. The unitary matrix $U_1$ that
diagonalizates a the potential matrix for the Rabi model is given in
(\ref{eigs}) by replacing the angle;
$\tan\left(2\theta\right)=2\sqrt{2}g_0x/\Omega$. From the identity
\begin{equation}\label{noncom}
U_1^\dagger pU_1=p-\sigma_y\hbar\partial\theta,
\end{equation}
where
\begin{equation}
\partial\theta\equiv\frac{\partial\theta}{\partial x}
\end{equation}
it is clear that the transformed Hamiltonian is not diagonal.
Explicitly we derive
\begin{equation}\label{adham}
\tilde{H}_{ad}=\frac{p^2}{2}+\frac{x^2}{2}+(\partial\theta)^2+\left[\begin{array}{cc}\lambda
& \displaystyle{\frac{\partial^2\theta-2i(\partial\theta)p}{2}} \\ \displaystyle{\frac{\partial^2\theta-2i(\partial\theta)p}{2}} & -\lambda\end{array}\right],
\end{equation}
where
\begin{equation}
\lambda=\sqrt{\frac{\Omega^2}{4}+2g_0^2x^2}
\end{equation}
and
\begin{equation}
\partial\theta=\frac{\sqrt{2}\Omega g_0}{\Omega^2+8g_0^2x^2},\hspace{1cm}\partial^2\theta=-\frac{16\sqrt{2}\Omega g_0^3}{\left(\Omega^2+8g_0^2x^2\right)^2}.
\end{equation}
The sizes of $\partial\theta$ and $\partial^2\theta$ measure the
amount of adiabaticity \cite{jonasad,com2}. From these we can draw
several conclusions; a large $\Omega$ favors an adiabatic evolution,
so does a large $x$. The first is the regular adiabatic dispersive
limit. The latter, however, is noticeably different from typical
adiabaticicity in the JC model, but non the less it is intuitive
since only close to the curve crossing is the adiabaticity assumed
to break down. It seems that also $\Omega\rightarrow0$ would give
adiabatic evolution, which, however, is false since this limit
describes the diabatic evolution. We recognize the {\it adiabatic
potentials}
\begin{equation}\label{rabipot}
V_{ad}^\pm(x)=\frac{x^2}{2}+(\partial\theta(x))^2\pm\lambda(x)
\end{equation}
and the internal {\it adiabatic basis states} as
$|\uparrow\rangle=\cos(\theta)|+\rangle+\sin(\theta)|-\rangle$ and
$|\downarrow\rangle=-\sin(\theta)|+\rangle+\cos(\theta)|-\rangle$.
The internal {\it diabatic basis states} are termed
$|u\rangle=\frac{1}{\sqrt{2}}(|+\rangle+|-\rangle)$ and
$|d\rangle=\frac{1}{\sqrt{2}}(|+\rangle-|-\rangle)$, with
corresponding {\it diabatic potentials}
$V_d^\pm(x)=V_h(x\pm\sqrt{2}g_0)$. Thus, a general {\it adiabatic}
or {\it diabatic state} is is given by
$\psi(x)|\uparrow(\downarrow)\rangle$ or $\psi(x)|u(d)\rangle$
respectively. Examples of the adibataic (solid) and diabatic
(dotted) potentials are presented in figure \ref{fig1}.

Labeling $H_{ad}^{(1)}=H_{ad}=U_1HU_1^\dagger$, the transformed
Hamiltonian $H_{ad}^{(1)}$ defines a new adiabatic diagonalization
matrix $U_2$ and so forth, leading to the expansion
$H_{ad}^{(n)}=\Big[U_nU_{n-1}\cdots U_2U_1\Big]H\Big[U_1^\dagger
U_2^\dagger\cdots U_{n-1}^\dagger U_n^\dagger\Big]$. We further note
that the rather general form of the adiabatic Hamiltonian
(\ref{adham}) shares great similarities to the JC one (\ref{conjc})
expressed in conjugate variables.

\section{Regarding the $p$-dependent coupling}\label{sec3}
The presence of the $p$-dependent term in the potential matrix of
the JC model (\ref{conjc}), makes its evolution, in some sense, more
complex (or less intuitive) than for its companion the Rabi model.
On the one hand, we know that the JC model is analytically solvable
and any physical quantity can in principle be obtained from certain
infinite sums. Non the less, the fact remains that the wave packet
dynamics is still more involved due to this term.

The analysis will be restricted to either initial Fock number, or
coherent states, which in $x$-representations read respectively
\begin{equation}\label{instate}
\begin{array}{l}
\displaystyle{\psi_n(x,0)=\frac{1}{\sqrt{2^nn!}}\left(\frac{1}{\pi}\right)^{1/4}\mathrm{H}_n(x)\mathrm{e}^{-\frac{x^2}{2}}},\\
\\
\displaystyle{\psi_\nu(x,0)=\left(\frac{1}{\pi}\right)^{1/4}\mathrm{e}^{-(\Im\nu)^2}\mathrm{e}^{-\frac{1}{2}\left(x-\sqrt{2}\nu\right)^2}},
\end{array}
\end{equation}
where $\mathrm{H}_n$ is the $n$th order Hermite polynomial and $\nu$
is the amplitude of the coherent state; $\bar{n}=|\nu|^2$. Thus, in
the case of a coherent state, the classical velocity
$v=\sqrt{2\bar{n}-g_0^2/2}$ at the crossing. For an initial Gaussian
(\ref{instate}) centered at $x_i=0$, corresponding to field vacuum,
one expects that the wave packet will split up and accelerate down
towards the two potential minima \cite{com3}. This is indeed what we
find for the Rabi model, while after all, this can not be the case
for the JC model since we know that for an initial Fock state, the
model exhibits Rabi oscillations. Thus, for the JC model the field
wave packet will either Rabi oscillate between the two states
$\psi_0(x)$ and $\psi_1(x)$ or remain in $\psi_0(x)$ throughout,
whether the initial atomic state is $|+\rangle$ or $|-\rangle$. This
constrain of the wave packet to the origin is a direct effect of the
$p$-dependent coupling, and holds for any Fock state $\psi_n(x)$ and
therefore also for Fock states extending over large ranges $x$. The
wave packet must be considered in its entirety as a single object,
or in other words the coherence extending over the whole wave packet
must be taken into account and one can not incoherently split up the
wave packet in separate individual pieces. From this fact,
understanding the influence of the $p$-dependent coupling becomes
more challenging. None the less, we have that for a $n$ photon Fock
state $\Delta p^2=\langle p^2\rangle-\langle p\rangle^2=n+1/2$ and
it is reasonable to expect that the "momentum" coupling will affect
the dynamics due to the large spread in $p$.

\begin{figure}[ht]
\begin{center}
\includegraphics[width=12cm]{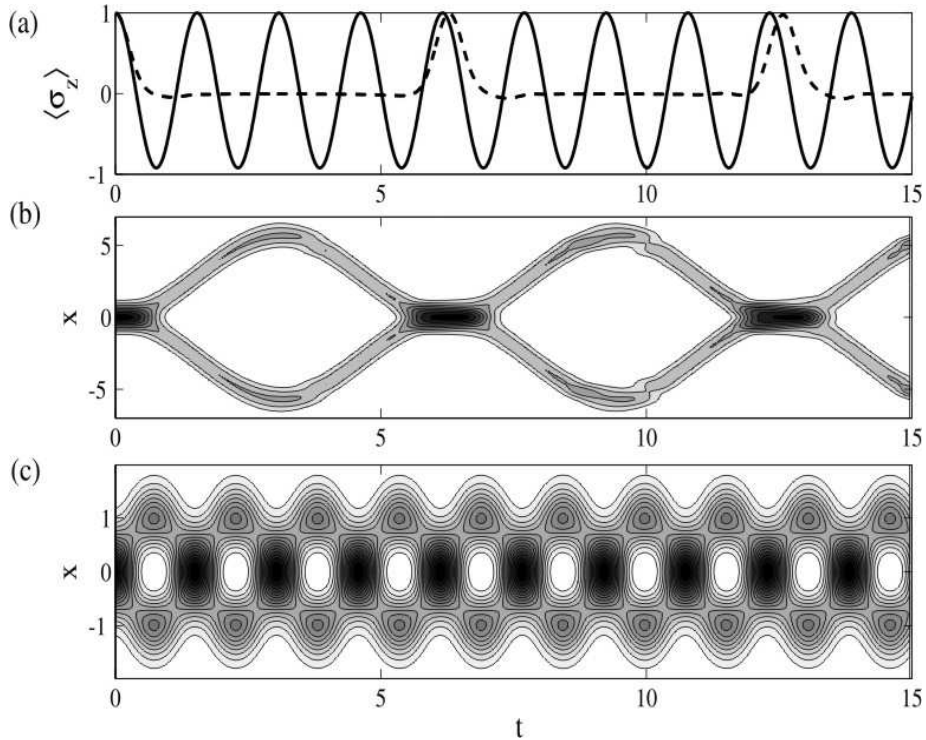}
\end{center}
\caption{\label{fig2} The effect of the $p$-dependent coupling in
the JC model. In the upper plot (a) we show the atomic inversion
$\langle\sigma_z\rangle$ between the two internal atomic states
$|\pm\rangle$ for the Rabi model (dashed) and the JC model (solid
line) and the initial state is here; field in vacuum and the atom
excited. The lower two plots presents the squared absolute
amplitudes $P(x,t)$ of the wave packet in the Rabi (b) and JC (c)
models. The parameters are $g_0=2$ and $\Omega=0.2$. }
\end{figure}

These conclusions are visualized in figure \ref{fig2}, which
displays the squared absolute amplitude of the wave packet in the
two models, together with the atomic inversion
$\langle\sigma_z\rangle$. We use the split operator method
\cite{split} to obtain the wave packet
\begin{equation}
|\Psi(x,t)\rangle=\left[\begin{array}{c}\psi_+(x,t)\\ \psi_-(x,t)\end{array}\right],
\end{equation}
and its squared absolute amplitude $P(x,t)=|\psi_+(x,t)|^2+|\psi_-(x,t)|^2$

\section{Collapse-revivals}\label{sec4}
\begin{figure}[ht]
\begin{center}
\includegraphics[width=12cm]{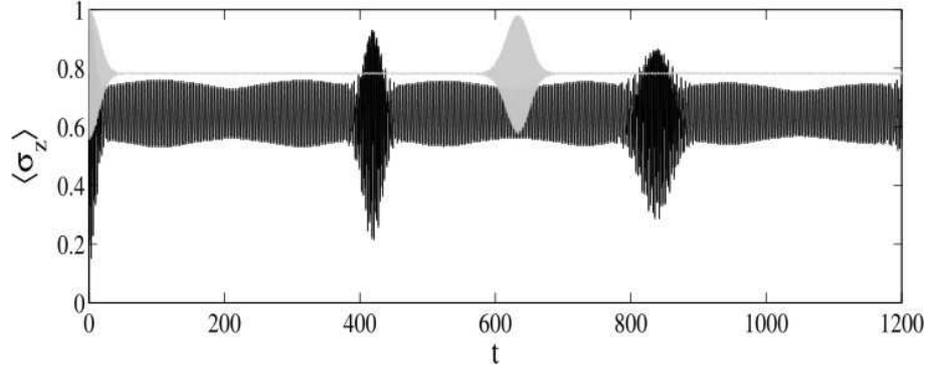}
\end{center}
\caption{\label{fig3} Inversion of the Rabi (black curve) and JC
model (gray curve) for an initial coherent state with amplitude
$\nu=7$, and dimensionless parameters $g_0=0.15$ and $\Omega=5$.  }
\end{figure}

\begin{figure}[ht]
\begin{center}
\includegraphics[width=12cm]{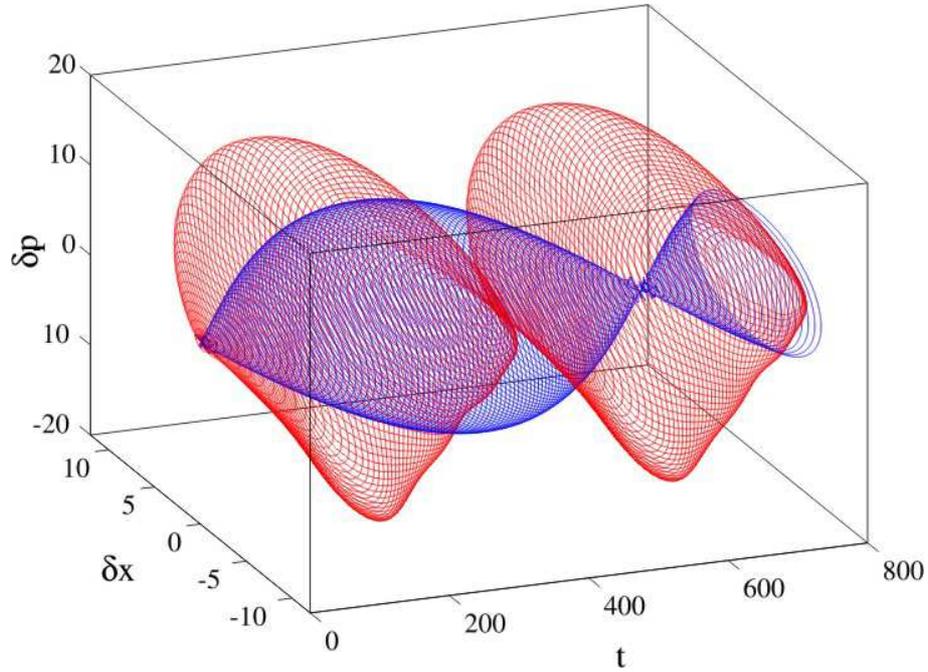}
\end{center}
\caption{\label{fig4} (Colour online) Inversion of the Rabi (red
curve) and JC model (blue curve) for an initial coherent state with
amplitude $\nu=7$, and dimensionless parameters $g_0=0.15$ and
$\Omega=5$.  }
\end{figure}

Collapses of physical variables are caused by dephasing between the
constitute terms making up the quantum state. While the collapses
are expected, more surprising is the phenomenon of revivals,
occurring when the terms return back in phase. Hence, the evolution
must be sufficiently coherent in order to be able to complentate the
rephasing. It is clear that revivals in the models considered in
this paper is a direct outcome of the quantized 'grainyness' of the
field and thus a novel quantum effect \cite{jcrev}. In other fields,
the existence of collapse-revivals has been greatly studied, and
probably the most significant contribution owes the one of wave
packet dynamics describing the vibrations in molecules \cite{wp}.
Here the wave packet evolves in a bound electronic potential, where
the discreteness derives from the vibrational eigen-modes in the
particular electronic molecular state. For a harmonic potential, a
wave packet bouncing back and forth in the potential, will reshape
after one period of oscillation, which defines the classical period
of motion. If the potential, however, is anharmonic the wave packet
will not fully reshape after one classical period, bringing about
the collapse. Depending on the degree of anharmonicity, the wave
packet may reshape at later times characterizing the revivals. For a
fairly localized excited wave packet, with average quantum number
$\bar{n}\gg1$, we assume $\Delta n/\bar{n}\ll1$ where $\Delta n$ is
the spread of quantum numbers. In this case we expand the
eigenenergies accordingly,
\begin{equation}\label{eigexp}
E(n)=E(n_0)+E'(n_0)(n-n_0)+\frac{E''(n_0)}{2}(n-n_0)^2+\frac{E'''(n_0)}{6}(n-n_0)^3...\,,
\end{equation}
where $E'(n_0)=(dE(n)/dn)_{n=n_0}$ and so on. The various terms
define different time scales according to
\begin{equation}\label{timescales}
\begin{array}{lll}
T_{cl}=\frac{2\pi}{|E'(n_0)|}, & T_{rev}=\frac{2\pi}{|E''(n_0)|}, & T_{sup}=\frac{2\pi}{|E'''(n_0)|},
\end{array}
\end{equation}
characterizing {\it classical}, {\it revival} and {\it superrevival}
times respectively. For a harmonic oscillator only the first of
these term is non-zero and identifies the classical period
$T_{cl}=2\pi/\omega$. We note that, assuming zero detuning,
$\Delta=0$, the JC energies expand as
$\sqrt{n}=\sqrt{\bar{n}}+\frac{1}{2\sqrt{\bar{n}}}(n-\bar{n})-\frac{1}{8\bar{n}^{3/2}}(n-\bar{n})^2+...\,$.
This is not, however, the standard way of deriving the revival times
in the JC model. Normally one solves for the time it takes for
consecutive Rabi frequencies to differ by a multiple of $2\pi$
\begin{equation}
\left(2\Omega_{\bar{n}+1}-2\Omega_{\bar{n}}\right)T'_{rev}=2\pi,
\end{equation}
where $\Omega_n=\sqrt{\frac{\Delta^2}{4}+g_0^2n}$. The derived
revival time is not identified with the previously defined $T_{rev}$
of equation (\ref{timescales}), but rather associates with the
classical time $T_{cl}$. In other words, $T_{cl}$ does {\it not}
correspond to the time for the constitute wave packets to bounce
back and forth in the potential. However, in the JC model, as well
as in the Rabi model, one has two internal degrees of freedom, and
the dynamics is obtained from both the internal wave packets and
their coupled motion. In particular, reshaping of the wave packet
means that both internal wave packets must reform simultaneously.
Thus, using the definitions (\ref{timescales}), $T_{cl}$ is
typically orders of magnitude larger in a multi-level system than
for an internal structure-less wave packet evolution. Another way of
picturing the phenomenon is that the constitute wave packets must
overlap in phase space in order to give rise to interference,
manifesting itself in the form of revivals. By introducing $\langle
x\rangle_\pm$ as the mean position of the two wave packets
$\psi_\pm(x,t)$, and similarly for the momentum $\langle
p\rangle_\pm$, revivals occur when $\delta x\equiv\langle
x\rangle_+-\langle x\rangle_-=0$ and $\delta p\equiv\langle
p\rangle_+-\langle p\rangle_-=0$ simultaneously. In figure
\ref{fig3} we show the atomic inversion and in figure \ref{fig4} the
quantities $\delta x$ and $\delta p$ as function of time $t$. In the
$x$-direction, both models oscillates between $\pm10$, while in the
momentum direction the Rabi model has $\langle p\rangle$ exceeding
10 at some occasions.

\begin{figure}[ht]
\begin{center}
\includegraphics[width=12cm]{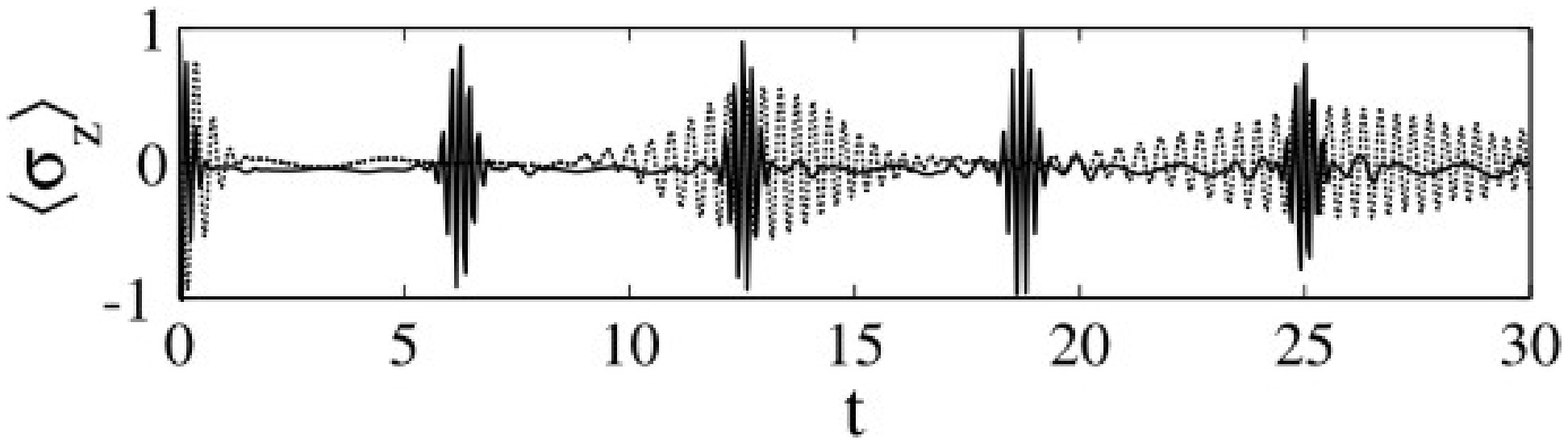}
\end{center}
\caption{\label{fig5} Inversion of the Rabi (solid line) and JC
model (dotted line) for an initial coherent state with amplitude
$\nu=4$, and dimensionless parameters $g_0=2$ and $\Omega=0.2$.  }
\end{figure}

\begin{figure}[ht]
\begin{center}
\includegraphics[width=12cm]{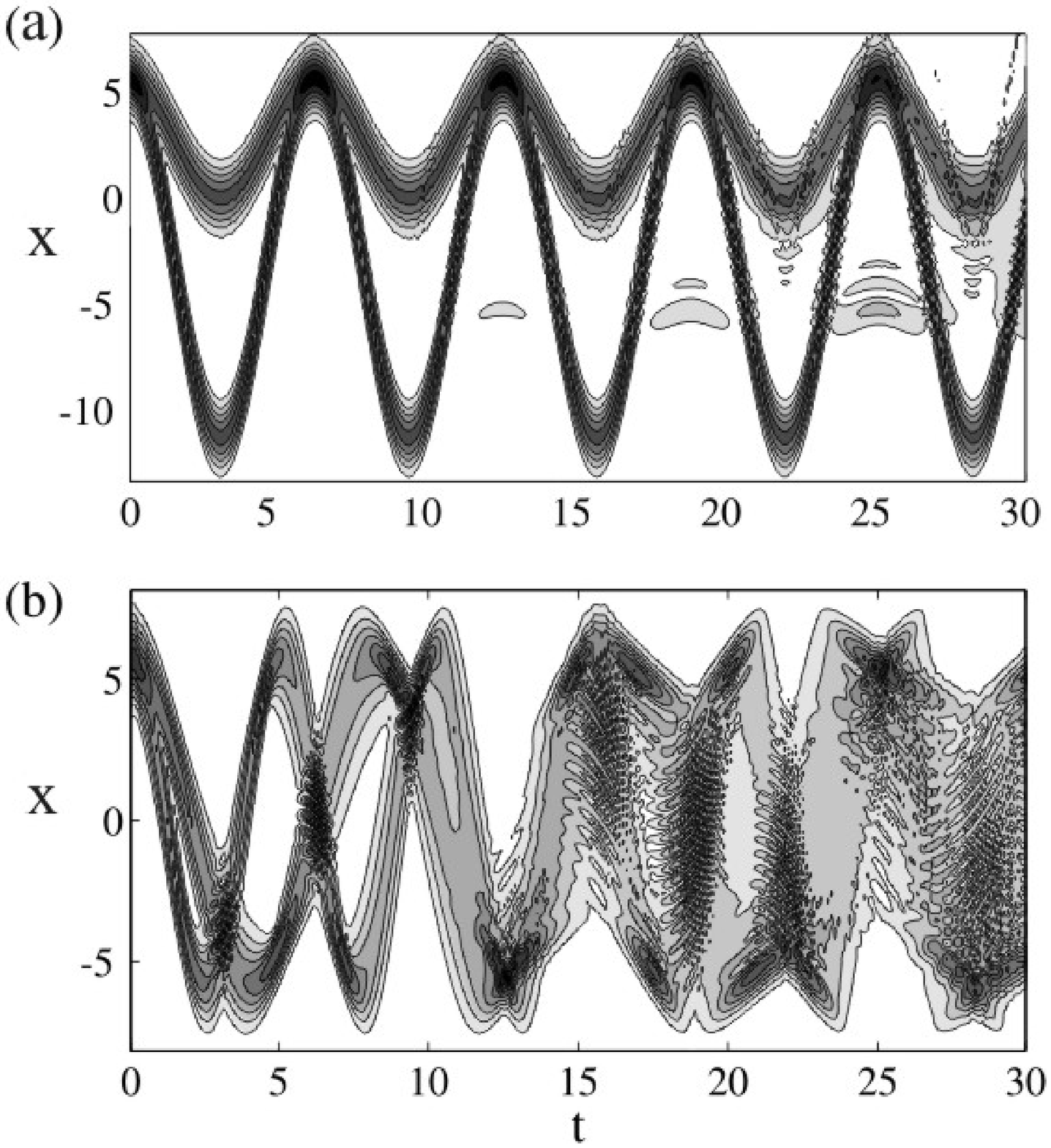}
\end{center}
\caption{\label{fig6} The squared amplitudes $P(x,t)$ for the Rabi
(a) and the JC (b) models corresponding to fig. \ref{fig5}. }
\end{figure}

As argued, the revivals obtained in the Rabi and JC models
correspond to the classical period $T_{cl}$ in equation
(\ref{timescales}), and not to $T_{rev}$ arising from the
anharmonities. Due to the internal two-level structure, the time
$T_{cl}$ can become rather long provided the proper adiabatic,
diabatic or mesobatic potentials differ in their corresponding
harmonic frequencies. However, if the frequencies almost coincide,
$T_{cl}$ does not need to be large as seen in the example of figures
\ref{fig5} (atomic inversion) and \ref{fig6} (wave packet
amplitudes). Here the parameters are such that the evolution is
approximately diabatic and the two Rabi diabatic potential curves
share almost identical curvature, causing the two wave packets to
oscillate with nearly the same classical frequencies. The JC model
shows more abstract dynamics, more reminiscent of the one seen in
figures \ref{fig3} and \ref{fig4}.

\section{Extension to multi-level internal structure}\label{sec5}
In \cite{oldwine}, the analysis was restricted to two-level atoms
interacting with a single cavity mode. The generalization to
multi-level atoms and/or multi-mode fields is straightforward.
However, the split operator method can only handle up to two
dimensions, limiting the numerics to at most two modes, while the
number of internal states may be as large as 20. However, other
approximate wave packet methods exists where the dimension might be
considerably larger  \cite{mch}. Here we only consider extensions in
the direction of multi-level internal states, and the many mode
situation will be dealt with in the future projects.

\subsection{Three-level $\Lambda$-atom}\label{ssec5a}
In the regular two-level atom model, losses of the excited level
often affects the dynamics in an undesired manner. This can be
circumvented by coupling two metastable "ground states" in a
three-level $\Lambda$-atom configuration, adiabatically eliminating
the excited state. However, the full three-level dynamics show
interesting features beyond the two-level atom situation
\cite{multi}. The model studied here is used mostly to present the
methods, as it can in fact be reduced into a two-level model.
Additionally, generalizing this model to non reducible three-level
systems is straightforward.

For simplicity we assume the two lower atomic states, $|g_1\rangle$
and $|g_2\rangle$, to be degenerate, and further that they dipole
couple to the excited state $|e\rangle$ through couplings
$\lambda_1$ and $\lambda_2$. The Hamiltonian (without the RWA) reads
\begin{equation}
H_\Lambda=\frac{p^2}{2}+\frac{x^2}{2}+\Omega\sigma_{ee}+\big[\lambda_1\left(\sigma_{g_1e}+\sigma_{eg_1}\right)+\lambda_2\left(\sigma_{g_2e}+\sigma_{eg_2}\right)\big]x,
\end{equation}
where $\sigma_{ij}=|i\rangle\langle j|$. It may be written in the
atomic bare basis $\left\{|g_1\rangle,|e\rangle,|g_2\rangle\right\}$
as
\begin{equation}
H_\Lambda=\frac{p^2}{2}+\frac{x^2}{2}+\left[\begin{array}{ccc} 0 &
\lambda_1x & 0 \\ \lambda_1x & \Omega & \lambda_2x \\ 0 & \lambda_2x
& 0\end{array}\right].
\end{equation}
Introducing the angles $\tan(\theta)=\lambda_1/\lambda_2$ and
$\tan(2\phi)=2\lambda_0x/\Omega$, with
$\lambda_0=\sqrt{\lambda_1^2+\lambda_2^2}$, the potential matrix is
diagonalized by
\begin{equation}
U_2=\left[\begin{array}{ccc} \sin(\phi)\sin(\theta) & \cos(\theta) &
\cos(\phi)\sin(\theta) \\
\cos(\phi) & 0 & -\sin(\phi) \\
\sin(\phi)\cos(\theta) & -\sin(\theta) &
\cos(\phi)\cos(\theta)\end{array}\right].
\end{equation}
As $\theta$ is $x$-independent we find
\begin{equation}
U_2pU_2^\dagger=p-iU_2\partial_\phi U_2^\dagger\partial_x\phi,
\end{equation}
and one derives that
\begin{equation}
U\partial_\phi U^\dagger=\left[\begin{array}{ccc}0 & 0 & 1 \\
0 & 1 & 0 \\ 1 & 0 & 0\end{array}\right].
\end{equation}
The adiabatic potentials are
\begin{equation}
V_\pm(x)=\frac{x^2}{2}+(\partial_x\phi)^2+\left(\frac{\Omega}{2}\pm\sqrt{\frac{\Omega^2}{4}+2\lambda_2^2x^2}\right),\hspace{1.2cm}V_0(x)=\frac{x^2}{2}+(\partial_x\phi)^2,
\end{equation}
where the adiabatic state corresponding to $V_0(x)$ is called the
{\it dark state} which becomes degenerate with the state of $V_-(x)$
at $x=0$. This degeneracy is lifted however, if a detuning is
assumed between the two ground states. We note that the two
adiabatic potentials $V_\pm(x)$ are similar to the ones of the Rabi
model (\ref{rabipot}). Actually, the absence of a second non-zero
diagonal element in the potential matrix gives the system a symmetry
such that it can be separated into a $2\times2$ problem. The unitary
transformation
\begin{equation}
U_3=\frac{1}{\lambda_0}\left[\begin{array}{ccc} \lambda_1 & 0 & \lambda_2\\
0 & \lambda_0 & 0 \\ \lambda_2 & 0 & -\lambda_2\end{array}\right]
\end{equation}
casts the potential matrix in the form
\begin{equation}
V'(x)=U_3V(x)U_3^\dagger=\left[\begin{array}{ccc} 0 & \lambda_0x & 0
\\ \lambda_0x & \Omega & 0\\ 0 & 0 & 0\end{array}\right].
\end{equation}

We may remark that like for the Rabi model, a semiclassical approach
of this $\Lambda$-system naturally leads to the solvable generalized
three-level Landau-Zener model \cite{3lz}. Wave packet propagation
methods in a related system has been considered in \cite{jonas2}.

\subsection{Dicke model}\label{ssec5b}
Extending the Rabi Hamiltonian to contain $N$ number of two-level
atoms gives the Dicke model \cite{dicke}
\begin{equation}
H_D=\frac{p^2}{2}+\frac{x^2}{2}+\Omega S_x+\frac{g_0}{\sqrt{N}}S_zx
\end{equation}
with the total spin observables $S_k=\sum_{i=1}^N\sigma_k^{(i)}$,
$k=x,z$ and $\sigma_k^{(i)}$ is the $k$'th Pauli matrix for atom
$i$. The scaling of the atom-field coupling by $\sqrt{N}$ is
inserted to assure a well defined thermodynamic limit as
$N\rightarrow\infty$. Operators acting on different atoms commutes,
so that the adiabatic atomic ground state reads
\begin{equation}
|E_0\rangle=\big\{\cos(\theta)|+\rangle-\sin(\theta)|-\rangle\big\}^{\otimes
N},
\end{equation}
where $\otimes N$ means direct product and as before
$\tan(2\theta)=2g_0x/\sqrt{N}\Omega$. Formally, the adiabatic
diagonalization procedure follows like for the single atom case,
replacing Pauli matrices by total spin operators. We unitary
transform the Hamiltonian by
\begin{equation}
U_4=\mathrm{e}^{-i\theta S_y}
\end{equation}
and the non-adiabatic corrections arise from equation (\ref{noncom})
substituting $\sigma_y$ by $S_y$. From the theory of angular
momentum we directly find the adiabatic potentials
\begin{equation}
V_{m_s}(x)=\frac{x^2}{2}+m_s\sqrt{\frac{\Omega^2}{4}+\frac{g_0^2}{N}x^2},\hspace{1,2cm}m_s=-N,-N+1,...,N-1,N.
\end{equation}
Here $m_s$ is the quantum number for the total spin in the
$z$-direction. It is convenient to introduce the {\it Dicke states}
$|s,m_s\rangle$ being eigenstates of the total spin
$S|s,m_s\rangle=s(s+1)|s,m_s\rangle$ and of the $z$-component
$S_z|s,m_s\rangle=m_s|s,m_s\rangle$. The adiabatic ground state thus
identifies with quantum numbers $s=N$ and $m_s=-N$, and applying the
rotated annihilation operator $\tilde{J}_+=U_4^\dagger J_+U_4$
repeated times to the ground state generates all the adiabatic
states $|N,m_s\rangle$. The states with lower $s$ quantum number can
be obtained by standard methods.

In the large $N$ limit, quantum fluctuations can in general be
regarded as small and one may linearize these terms. The
Holstein-Primakoff representation of the spin operators \cite{hp}
has turned out to be an efficient approach in analyzing the Dicke
model \cite{eb1,eb2}. The spin operators are expressed in boson
operators accordingly;
\begin{equation}
S_z=b^\dagger
b.\frac{N}{2},\hspace{1cm}S_+=b^\dagger\sqrt{N-b^\dagger
b},\hspace{1cm}S_-=S_+^\dagger.
\end{equation}
Before linearizing, we note from figure \ref{fig1} that the low
lying adiabatic potentials (we assume cold atoms and therefore a low
temperature) either have one or two global minima, and this fact
should be taken into account for when expanding in $N^{-1}$. This is
directly related to the presence of a quantum phase transition, with
critical coupling $g_0^{(c)}=\sqrt{\Omega}/2$, between the {\it
normal phase} of a vacuum field and all atoms deexcited and the {\it
superradient phase} of a macroscopic field and atomic excitation
\cite{dickept1,dickept2}. Due to this it is convenient to coherently
shift the boson operators as \cite{eb1,eb2}
\begin{equation}
a\rightarrow c+\alpha_s,\hspace{1.2cm}b\rightarrow d+\beta_s,
\end{equation}
where
\begin{equation}
\alpha_s=g_0\sqrt{N(1-\mu^2)},\hspace{1.2cm}\beta_s=\sqrt{\frac{N}{2}(1-\mu)}
\end{equation}
and
\begin{equation}
\mu=\left\{\begin{array}{llll} \left(\frac{g_0^{(c)}}{g_0^2}\right)^2,& & g_0>g_0^{(c)}\\
1, & & g_0\leq g_0^{(c)}\end{array}\right..
\end{equation}
Thus, we note that in the case of a single global minimum the shifts
vanish. Here $c$ and $d$ represents quantum fluctuations around the
classical values $\alpha_s$ and $\beta_s$. The resulting expanded
Dicke Hamiltonian becomes \cite{eb1,eb2}
\begin{equation}
H_D'=c^\dagger c+\frac{\Omega(1+\mu)}{2\mu}d^\dagger
d+\frac{\Omega(1-\mu)(3+\mu)}{8\mu(1+\mu)}(d^\dagger+d)^2+g_0\mu\sqrt{\frac{2}{1+\mu}}(c^\dagger+c)(d^\dagger+d).
\end{equation}
For $\mu=1$ (the ground adiabatic potential possess only a single
minimum) the Hamiltonian becomes bi-linear and one may decouple the
two boson modes into two disconnected harmonic oscillators.

Using the Holstein-Primakoff representation we turn the fermionic
degrees of freedom into a single bosonic degree of freedom. The cost
of reducing the $2^N$ potential curves to a single one is that the
wave packet now evolves in 2-D rather than 1-D. Non the less, wave
packet propagation in 2-D is easily performed with the split
operator method, and will be consider in future works. Another
algebraic method that could be considered is the {\it "Schwinger's
oscillator model of angular momentum"} \cite{schwinger}, in which,
however, two bosonic degrees of freedom is introduced instead of the
fermionic subsystem.

\section{Conclusion}\label{sec6}
The method of wave packet propagation, in which the cavity fields
quadrature operators serve as conjugate variables, has been
explored. It has been applied to the seminal JC model and its
companion, the Rabi model, which is related to the JC one by the
RWA. Various bases, different form the conventional bare and dressed
bases, and their corresponding potential curves were consider. This
new numerical approach is more commonly used in chemical and
molecular physics, were it has been utilized for more than three
decades \cite{heller}. The effect of the RWA was discussed in this
representation, and rather unexpected phenomena of the wave packet
evolution appear once the RWA has been assumed.

The main part of the analysis has been devoted to the
collapse-revival effect of these models. Typically it is studied
using algebraic methods, while we examine it from the shapes of the
coupled potential curves. The terminology of collapse-revivals in
wave packet models is not the same as the one for the JC model and
we sort this out. The internal two-level structure of the Rabi and
JC models may cause very long characteristic time scales, compared
to the ones of a wave packet in a single anharmonic potential.

Finally we have sketched how the method is extended to systems with
more internal degrees of freedom, and here especially to the
three-level $\Lambda$-atom and the Dicke model. For the Dicke model,
we show how one may transform the internal degrees of freedom into a
single "external" degree of freedom using the Holstein-Primakoff
representation. More thorough research on multi-level and multi-mode
systems are left for future works.

\ack

The author acknowledges support from the Swedish
government/Vetenskapsr{\aa}det and the European Commission (EMALI,
MRTN-CT-2006-035369; SCALA, Contract No. 015714).


\begin{thebibliography}{9}
\bibitem{com1} The effective atom-field coupling scales as $g_0\sqrt{n}$ and therefore this conditiond modifies to $g_0\sqrt{\bar{n}}<\Omega$, where $\bar{n}$ is the average number of photons.
\bibitem{com2}Note that, apart from the derivative terms $\partial\theta$ and $\partial^2\theta$, the ``momentum'' $p$ enters in the off-diagonal elements of (\ref{adham}) in agreement with the conclusions of subsection \ref{ssec2b}.
\bibitem{com3} Here we have assumed paramteres such that the lower adiabatic potential pocesses two minima.
\end{thebibliography}
\end{document}